\documentclass[
    aps,
    prb,
    twocolumn,
    floatfix,nofootinbib,
    superscriptaddress
]{revtex4-2}

\usepackage{subfigure}
\usepackage[utf8]{inputenc}
\usepackage{times}
\usepackage[breaklinks]{hyperref}
\usepackage{color}
\usepackage{siunitx}
\usepackage{booktabs}
\usepackage[version=4]{mhchem}
\usepackage{booktabs}

\usepackage{amssymb} 
\usepackage{amsmath}
\usepackage{amsthm}
\usepackage{braket}
\usepackage{stmaryrd}
\usepackage{mathtools}

\newcommand{\pgi}{Peter Gr\"unberg Institut and Institute for Advanced Simulation,
Forschungszentrum J\"ulich and JARA, 52425 J\"ulich, Germany}

\newcommand{\aachen}{Department of Physics, RWTH Aachen University, 52056 Aachen, Germany}

\newcommand{\mainz}{Institute of Physics, Johannes Gutenberg University Mainz, 55099 Mainz, Germany}

\usepackage{tikz}
\usepackage{tikz-cd}
\usetikzlibrary{calc}
\usetikzlibrary{decorations.markings}
\usepackage{xcolor}
\usepackage{graphicx}

\makeatletter
\renewcommand\@biblabel[1]{#1.}
\makeatother
\begin{document}

\setcounter{secnumdepth}{2} 


\title{
Photocurrents of charge and spin in single-layer Fe$_3$GeTe$_2$}

\author{M. Merte}
	\thanks{m.merte@fz-juelich.de}
    \affiliation{\pgi}
    \affiliation{\aachen}
    \affiliation{\mainz}

\author{F. Freimuth}
\thanks{f.freimuth@fz-juelich.de}
\affiliation{\mainz}
    \affiliation{\pgi}

\author{T. Adamantopoulos}

    \affiliation{\pgi}

\author{D. Go}

\affiliation{\mainz}
    \affiliation{\pgi}
    
    \author{T.G. Saunderson}

\affiliation{\mainz}
    \affiliation{\pgi}
    
    \author{M. Kl\"aui}

\affiliation{\mainz}
  
  \author{L. Plucinski}
  \affiliation{\pgi}
    
 \author{O. Gomonay}

\affiliation{\mainz}





\author{S. Bl\"ugel}

    \affiliation{\pgi}
    

    
\author{Y. Mokrousov}
\thanks{y.mokrousov@fz-juelich.de}
    \affiliation{\pgi}
    \affiliation{\mainz}

\date{\today}

\begin{abstract}
In the realm of two-dimensional materials magnetic and transport properties of a unique representative $-$ Fe$_3$GeTe$_2$ $-$ attract ever increasing attention.
Here, we use a developed first-principles method for calculating laser-induced response to study the emergence of photo-induced currents of charge and spin in single-layer Fe$_3$GeTe$_2$, which are of second order in the electric field. 
We provide a symmetry analysis of the emergent photocurrents in the system finding it to  be in excellent agreement with {\it ab-initio} calculations. 
We analyse the magnitude and behavior of the charge photocurrents with respect to disorder strength, frequency and band filling. Remarkably, not only do we find a large charge current response, but also predict that Fe$_3$GeTe$_2$ can serve as a source of significant laser-induced spin-currents, which makes this material as a promising platform for various applications in optospintronics.  

\end{abstract}

\maketitle






\date{\today}


%
%
%
%
%
%
%
%
%
%
%
%
%
%
%

{\it Introduction.} 
The tremendous progress in realization of robust two-dimensional (2D) magnetism in van der Waals materials~\cite{Huang_2017,Gong_2019,Gong_2017,Gibertini_2019} moves various properties of 2D magnetic materials into the focus of intense attention. Among the latter, layered Fe$_3$GeTe$_2$ (FGT) is one of the most prolific candidates for potential 2D magnetic applications, as it is one of the few compounds among 2D materials that exhibits strong out of plane magnetocrystalline anisotropy, has one of the largest Curie temperatures among 2D materials~\cite{Verchenko_2015}, and provides a playground for realization of complex spin textures~\cite{Ding_2020,Wu_2020,Park_2021}. Current-induced switching of magnetization in FGT has been achieved~\cite{Alghamdi_2019}, and it was argued that intrinsic bulk-like spin-orbit torques, arising without a need for an interface, can be very large in this material~\cite{Johansen_2019,arXiv:2107.09420}. Moreover,  it was predicted that FGT displays very prominent Kerr and Faraday effects, with magneto-optical properties being qualitatively similar when going from bulk to the single-layer limit owing to the weak coupling among the layers~\cite{ arXiv:2108.02926,arXiv:2012.04285v1}. These findings motivate an extensive further exploration of intrinsic properties of FGT, especially in the area of its magnetic response to  electromagnetic fields.  

On this front,  the properties of laser ignited charge currents are currently studied very intensively in interfacial systems and 2D materials, since they mediate THz radiation~\cite{Kampfrath_2013,Dhillon_2017,Vitiello_2019,PapaioannouBeigang} and carry important information about intrinsic characteristics of the system~\cite{wahada_2021,Burch_2018,PapaioannouBeigang,huisman_2016}. 
While it is known that even in nonmagnetic non-centrosymmetric materials such as  semiconductors~\cite{Sipe_2000,Mu_2021}, quantum wells~\cite{Sherman_2005}, 
graphene~\cite{Yin_2019} and organic-inorganic halides~\cite{Tyznik_2021} light can give rise to spin currents, following the initial suggestion of enhanced surface spin photocurrents in magnetic  systems~\cite{freimuth_2021}, the physics of laser-driven spin currents in 2D magnetic materials has started to attract attention as well~\cite{Xu_2021,Xiao_2021}.  
In this work we study the properties of laser-induced in-plane charge and spin currents in a single-layer FGT from first principles. For this, we employ an {\it ab-initio} implementation of the expressions for photocurrents of spin and charge that we derived recently~\cite{freimuth_2021}, which work equally well for insulating and metallic systems of any given complexity, and allow for considering the effect of disorder. We compute the charge and spin current response of FGT for different degree of disorder, and analyze it as a function of frequency and band filling. Our results provide an important reference point for exploring future optospintronics applications of this exciting material.
{\it Method.} In order to compute the photocurrents in the system arising as a response to a continuous laser pulse of frequency $\omega$, we employ an expression for the second order photocurrent density which was previously derived by us using Keldysh formalism
~\cite{freimuth_2021}:
\begin{equation}
    J_{i}=\frac{a_{0}^{2} e \epsilon_{0}}{2\hbar }\left(\frac{\varepsilon_{\mathrm{H}}}{ \hbar \omega}\right)^{2} \operatorname{Im} \sum_{j k} E_{j} E_{k}^{*} \varphi_{i j k},
    \label{eq:photocurrent}
\end{equation}
where $a_0$ is the Bohr's radius, $e$ is the elementary charge, $\hbar$ is the reduced Planck constant, $\varepsilon_H=e^2/(4\pi \epsilon_0 a_0)$ is the Hartree energy, and $E_i$ is the $i$'th component of the complex field amplitude of the pulse. 
The quantity $\varphi_{ijk}$
is defined as the following energy integral~\cite{freimuth_2021}:
\begin{equation}
    \begin{aligned}
\varphi_{i j k}=& \frac{2}{a_{0} \mathcal{E}_{\mathrm{H}}} \int \frac{\mathrm{d}^{2} k}{(2 \pi)^{2}} \int \mathrm{d} \mathcal{E} \operatorname{Tr}[\\
& f(\mathcal{E}) v_{i} G_{\boldsymbol{k}}^{\mathrm{R}}(\mathcal{E}) v_{j} G_{\boldsymbol{k}}^{\mathrm{R}}(\mathcal{E}-\hbar \omega) v_{k} G_{\boldsymbol{k}}^{\mathrm{R}}(\mathcal{E}) \\
-& f(\mathcal{E}) v_{i} G_{\boldsymbol{k}}^{\mathrm{R}}(\mathcal{E}) v_{j} G_{\boldsymbol{k}}^{\mathrm{R}}(\mathcal{E}-\hbar \omega) v_{k} G_{\boldsymbol{k}}^{\mathrm{A}}(\mathcal{E}) \\
+& f(\mathcal{E}) v_{i} G_{\boldsymbol{k}}^{\mathrm{R}}(\mathcal{E}) v_{k} G_{\boldsymbol{k}}^{\mathrm{R}}(\mathcal{E}+\hbar \omega) v_{j} G_{\boldsymbol{k}}^{\mathrm{R}}(\mathcal{E}) \\
-& f(\mathcal{E}) v_{i} G_{\boldsymbol{k}}^{\mathrm{R}}(\mathcal{E}) v_{k} G_{\boldsymbol{k}}^{\mathrm{R}}(\mathcal{E}+\hbar \omega) v_{j} G_{\boldsymbol{k}}^{\mathrm{A}}(\mathcal{E}) \\
+& f(\mathcal{E}-\hbar \omega) v_{i} G_{\boldsymbol{k}}^{\mathrm{R}}(\mathcal{E}) v_{j} G_{\boldsymbol{k}}^{\mathrm{R}}(\mathcal{E}-\hbar \omega) v_{k} G_{\boldsymbol{k}}^{\mathrm{A}}(\mathcal{E}) \\
+&\left.f(\mathcal{E}+\hbar \omega) v_{i} G_{\boldsymbol{k}}^{\mathrm{R}}(\mathcal{E}) v_{k} G_{\boldsymbol{k}}^{\mathrm{R}}(\mathcal{E}+\hbar \omega) v_{j} G_{\boldsymbol{k}}^{\mathrm{A}}(\mathcal{E})\right]
\end{aligned}
\label{eq:conductivity}
\end{equation}
where $v_{i}$ is the $i$'th component of the velocity operator, $f(\mathcal{E})$ is the Fermi-Dirac distribution function and $G_{\boldsymbol{k}}^{\mathrm{R/A}}$ are equilibrium retarded and advanced Greens functions respectively. In order to compute the spin photocurrent $Q_{si}$ propagating in direction $i$ and polarized along axis $s$, we replace the first of the velocity operators $v_i$ appearing in the expression above, with the operator of the spin velocity $\{v_i,\sigma_s\}$, and change the prefactor $a_0^2e \epsilon_0 /2\hbar $ in Eq.~\ref{eq:photocurrent} to $-a_0^2 \epsilon_0 /8$. 

In this work, we model the effect of disorder by adapting a model of constant lifetime broadening of the states $\Gamma$ which results in the following expressions: $G_{\boldsymbol{k}}^{\mathrm{R}}(\mathcal{E})=\hbar \sum_{n} \frac{|\boldsymbol{k} n\rangle\langle \boldsymbol{k} n|}{\mathcal{E}-\mathcal{E}_{\boldsymbol{k} n}+i \Gamma}$ and  $ G_{\boldsymbol{k}}^{\mathrm{A}}(\mathcal{E})= [ G_{\boldsymbol{k}}^{\mathrm{R}}(\mathcal{E}) ]^\dagger$~\cite{freimuth_2021}, where the energy of the state  $\Ket{\boldsymbol{k}n}$ in  band $n$ with a Bloch vector $\boldsymbol{k}$ is $\mathcal{E}_{\boldsymbol{k} n}$.
 To compute the photocurrent the integrals in Eq.~\eqref{eq:conductivity} have to be evaluated. At zero temperature the Fermi distribution becomes a step function which allows one to perform the energy integration analytically
 Ref.~\cite{freimuth_2016}. The numerical evaluation is preformed within the basis of maximally localized Wannier functions, and the  Brillouin zone integration  is performed numerically by employing the efficient technique of Wannier interpolation~\cite{code_w90,code_fleurWann,code_w90_ahc}. A more detailed description of the methodology will be published elsewhere. Throughout this work we assume an intensity of the light of 10\,GW/cm$^2$, which corresponds to typical values of the fluence of the order of 0.5\,mJ/cm$^2$ for a 50\,fs laser pulse~\cite{huisman_2016}.

\begin{figure}[t!]
\begin{center}
\rotatebox{0}{\includegraphics [width=0.92\columnwidth]{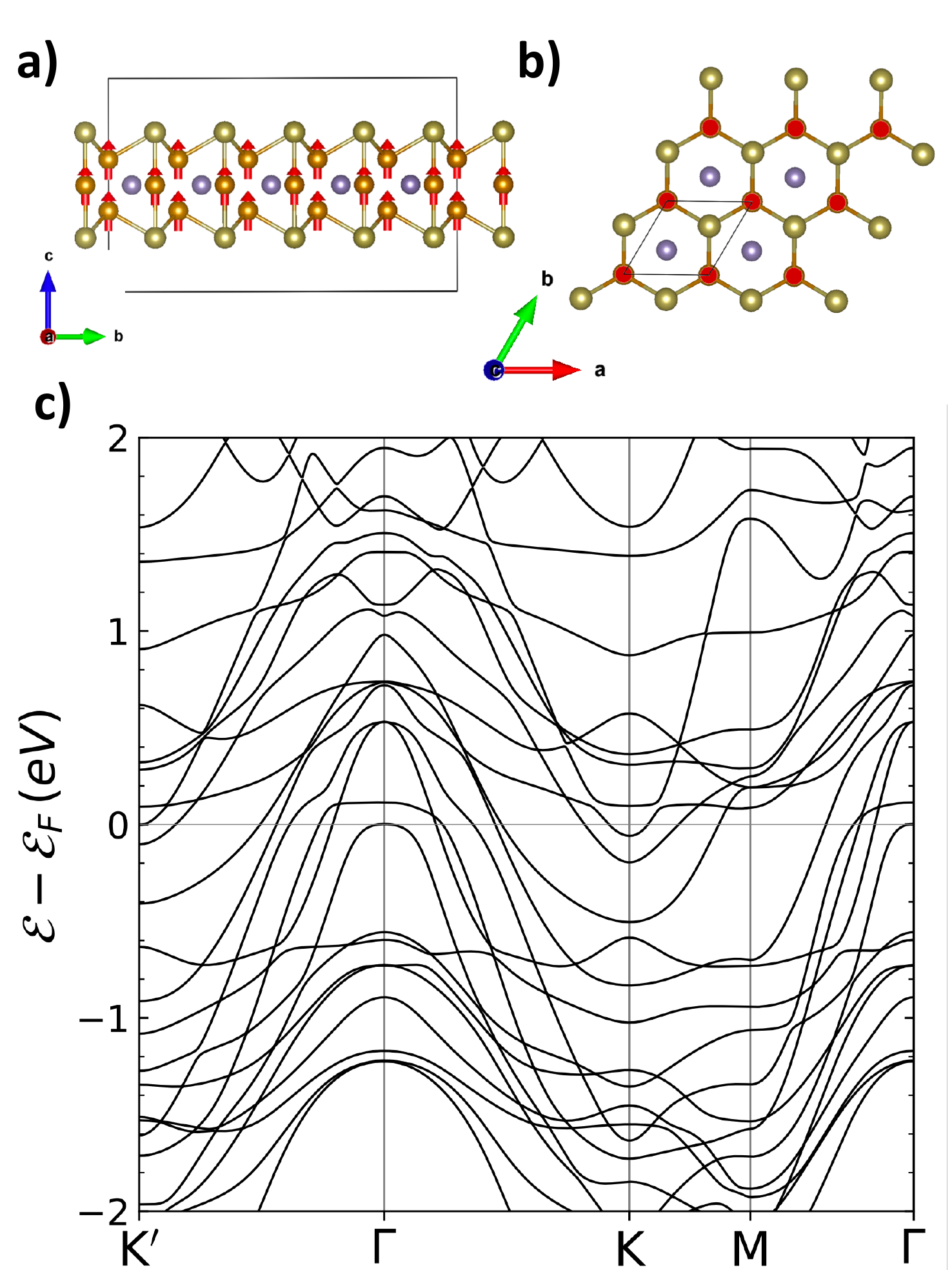}}
\end{center}
\caption{Geometry and electronic strcuture of single-layer Fe$_3$GeTe$_2$ (FGT). (a) Side view of FGT layer with red arrows indicating magnetic moments on Fe atoms directed out of plane. (b) Top view of a 2$\times$2 unit cell of single-layer FGT. In (a-b) blue spheres, green spheres and red spheres stand for Ge, Te and Fe atoms, respectively. (c) Bandstructure of ferromagnetic FGT layer in a $[-2,+2]$\,eV energy window around the Fermi energy.}
\label{fig:FGT_001_struct_bands}
\end{figure}

{\it Computational Details.} 
The considered structure of single-layer FGT is shown in Fig.~1(a-b). Our calculations of single-layer Fe$_3$GeTe$_2$ with point group D$_{3h}$ predict that in agreement with experiments the FGT layer  exhibits a ferromagnetic ground state with the easy axis pointing out of the plane when grown at the lattice constant of bulk Fe$_3$GeTe$_2$ ~\cite{Huang_2017}.  
The electronic structure of the system was calculated including the effect of spin-orbit coupling (SOC) with the film version of the \texttt{FLEUR} code~\cite{fleur}. The in-plane lattice constant was set to 
$a=7.542$\,a.u.
For self-consistent calculations we used a plane-wave cutoff of 
$5.0$\,a.u.$^{-1}$ and the total of 576 $k$-points in the two-dimensional Brillouin zone. 
The muffin-tin radii for Fe, Ge, Te were set to 2.08\,a.u., 2.19\,a.u., and 2.40\,a.u., respectively. The nonrelativistic PBE~\cite{pbe} exchange-correlation functional was used. The computed bandstructure of the system, shown in Fig.~\ref{fig:FGT_001_struct_bands}, reflects the complex orbital interplay around the Fermi energy $\mathcal{E}_F$ typical of FGT. The impact of low symmetry of FGT on the electronic structure can be clearly observed along the ${\rm K}'-\Gamma-{\rm K}$ path in Fig.~\ref{fig:FGT_001_struct_bands}, and it ultimately gives rise to non-vanishing complex non-linear response.  After converging the electronic structure we extracted 48 maximally-localized Wannier functions by using Fe $d$- and Ge, Te $p$-orbitals as initial projections, since these orbitals dominate the bandstructure in a wide energy window around $E_F$. 
Based on the tight-binding Wannier Hamiltonian constructed from the Wannier functions we computed the photocurrents on a  $2000\times2000$ interpolation $k$-mesh, which provides well converged results for the lifetime broadening larger than $\Gamma=25$\,meV.





{\it Symmetry analysis.} 
In order to investigate which constraints are imposed on the 
photocurrents by the crystal symmetry we expand the photocurrent
up to the first order in the magnetization as follows:
\begin{equation}
J_i=2\sigma_{ijk}E_jE_k^{*} = \chi_{ijk}E_j E_k^{*}+\chi_{ijkl}E_j E_k^{*} M_l,
\label{eq:symmetry}
\end{equation}
where $\sigma_{ijk}$ is the photoconductivity tensor, $\chi_{ijk}$  is a polar tensor of rank 3 and
$\chi_{ijkl}$ is an axial tensor of rank 4.
We introduce the notation for the basis tensors
$\delta^{(nopq)}_{ijkl}=\delta_{in}\delta_{jo}\delta_{kp}\delta_{lq}:=
\langle nopq \rangle$,
which allows us to list the tensors that are permitted by symmetry
in a compact form.
Only one polar tensor of rank 3 is allowed by symmetry, namely
$\chi^{(p,1)}_{ijk}=\delta^{(111)}_{ijk}-\delta^{(221)}_{ijk}-\delta^{(212)}_{ijk}-\delta^{(122)}_{ijk}$,
while four axial tensors of rank 4
are consistent with the symmetry:
$\chi^{(a,1)}_{ijkl}=-\delta^{(3211)}_{ijkl}-\delta^{(3121)}_{ijkl}-\delta^{(3112)}_{ijkl}+\delta^{(3222)}_{ijkl}$, $\chi^{(a,2)}_{ijkl}=-\delta^{(2113)}_{ijkl}-\delta^{(1213)}_{ijkl}-\delta^{(1123)}_{ijkl}+\delta^{(2223)}_{ijkl}$, $\chi^{(a,3)}_{ijkl}=-\delta^{(2131)}_{ijkl}-\delta^{(1231)}_{ijkl}-\delta^{(1132)}_{ijkl}+\delta^{(2232)}_{ijkl}$, and $\chi^{(a,4)}_{ijkl}=\delta^{(2311)}_{ijkl}+\delta^{(1321)}_{ijkl}+\delta^{(1312)}_{ijkl}-\delta^{(2322)}_{ijkl}$.
%
We do not need to consider $\chi^{(a,1)}$
due to the
2D character of FGT, where out-of-plane photocurrents are not of
interest. Additionally, we may ignore $\chi^{(a,3)}$
and $\chi^{(a,4)}$
because we choose the magnetization 
direction along the $z$ axis. Considering the remaining tensors, expressions for $\chi^{(p,1)}$ and $\chi^{(a,2)}$
predict vanishing photocurrents for the circularly polarized pulses, while the sign of $J_i$ switches upon the change in the direction of light from $x$ to $y$ for linearly polarized light.
For the spin current we perform a similar symmetry analysis by expanding it as
\begin{equation}
Q_{si}=\chi_{sijk}E_j E_k^{*}+\chi_{sijkl}E_j E_k^{*} M_l,
\end{equation}
where $\chi_{sijk}$ is an axial tensor of rank 4 while
$\chi_{sijkl}$ is a polar tensor of rank 5. 
The
expressions for axial tensors of rank 4 are listed above, 
and due to the 2D character of FGT, we need to consider only
$\chi^{(a,1)}$ 
in the expansion of $Q_{si}$ (note that in the expansion of $J_i$ we excluded $\chi^{(a,1)}$). 
 There are fifteen polar tensors of rank 5 for the point group of FGT.
As discussed above we need to consider only tensors that predict an effect for magnetization along the $z$ axis. Additionally, we consider only tensors where the indices
$i,j,k$ are all different from $z$ because of the 2D character of FGT. It turns out that out of the 15 polar tensors only a single one satisfies these requirements. It is given by
$\chi^{(p,1)}_{sijkl}=\delta^{(31113)}_{sijkl}-\delta^{(32213)}_{sijkl}-\delta^{(32123)}_{sijkl}-\delta^{(31223)}_{sijkl}$. As a result, by symmetry only $z$-polarized spin currents  are allowed, with in-plane properties of spin currents identical to those of charge currents.

 \begin{figure}[t!]
\begin{center}
\rotatebox{0}{\includegraphics 
[width=0.95\columnwidth]{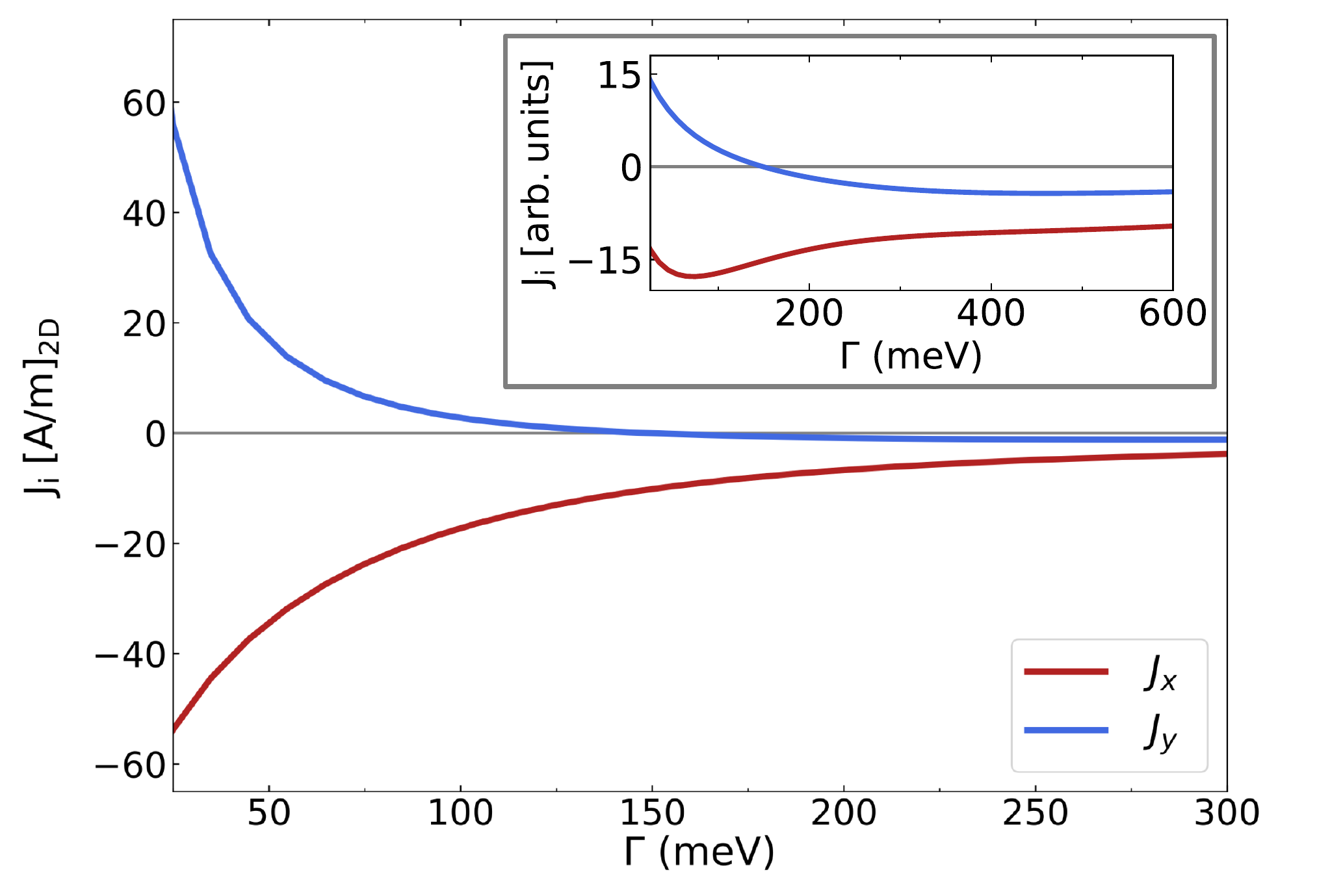}}
\end{center}
\caption{Dependence of photocurrents on the band broadening $\Gamma$. Shown are the photocurrent components $J_x$ (red solid line) and $J_y$ (blue dashed line) in response to the light pulse of frequency $\hbar \omega=1.55$\,eV and intensity of 10\,GW/cm$^2$
polarized along the $x$-axis.  
The inset depicts the corresponding photocurrent components multiplied with a factor $\Gamma$.
}
\label{fig:FGT_001_linPol_nk8k}
\end{figure}

{\it Results.} Overall, within the accuracy of the calculations, the symmetry properties of computed photocurrents are fully consistent with the symmetry analysis presented above, and thus in the following we focus only on $J_x$ and $J_y$ component of the charge current, and $Q_{zx}$, $Q_{zy}$ components of the spin current, arising in response to a laser pulse polarized along $x$-direction. First, we calculate and present in Fig.~\ref{fig:FGT_001_linPol_nk8k} the dependence of the charge photocurrents on the quasi-particle lifetime as quantified by parameter $\Gamma$ for the same light frequency of  1.55\,eV as used in Ref.~\cite{freimuth_2021}. 
For small values of $\Gamma$ between 25 $-$ 50\,meV the predicted magnitude of the photocurrent lies in the vicinity of 50\,A/m, which is roughly one order of magnitude larger than the magnitude of photocurrents emerging in the magnetic Rashba model with the Rashba strength of 100\,meV and similar parameters of the pulse and degree of disorder, as predicted in Ref.~\cite{freimuth_2021}. This highlights FGT as a source of strong intrinsic  photocurrents which originate in the symmetry properties of this material, and which will emerge in addition to interfacial currents when deposited on a substrate.  

\begin{figure}[t!]
\begin{center}
\rotatebox{0}{\includegraphics [width=0.95\columnwidth]{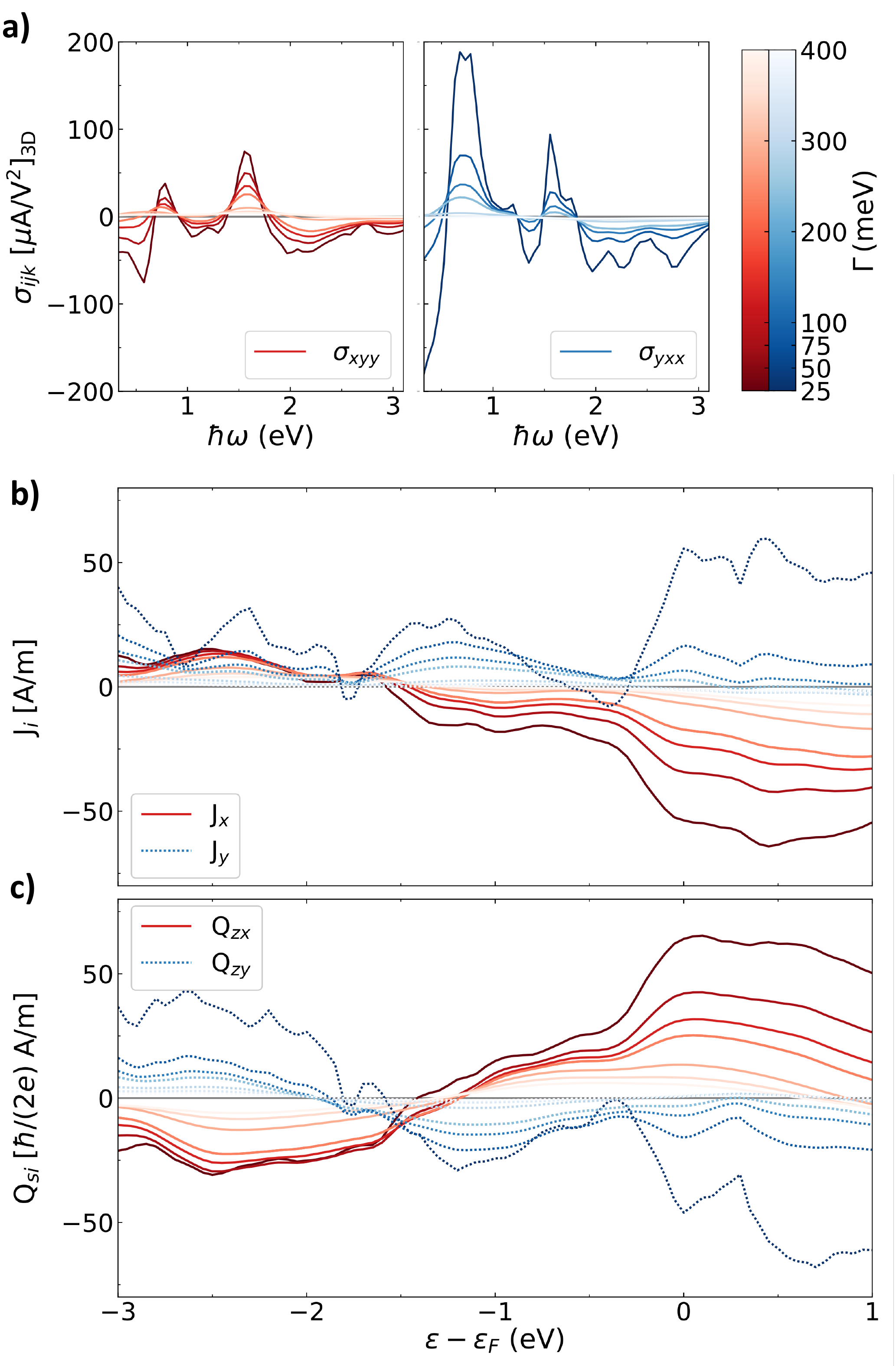}}
\end{center}
\caption{Spectral properties of the photocurrents. (a) Frequency dependence of the conductivity tensor components $\sigma_{xyy}$ and $\sigma_{yxx}$ evaluated for the true position of the Fermi energy in FGT layer for different values of the broadening. (b) Dependence of the photocurrents $J_x$ and $J_y$ on the position of the Fermi level and disorder strength (fading color of the lines). (c) Same as in (a) for spin photocurrents $Q_{xz}$ and $Q_{zy}$. 
In (b-c) the light pulse of frequency $\hbar \omega=1.55$\,eV and intensity of 10\,GW/cm$^2$
polarized along the $x$-axis was used. In all plots $\Gamma$ is indicated with the color scale.}
\label{fig:FGT_001_110_pol}
\end{figure}

The magnitude of the photocurrents rapidly decreases as the amount of disorder increases in the system, and the signal decreases ten-fold by going to the values of $\Gamma$ of about 300\,meV, with $J_y$ changing sign at the disorder strength of about 150\,meV. Generally, the overall functional dependence of the photocurrents on lifetime currently presents a subject of debates, as the non-linear nature of the effect makes it extremely difficult to disentangle various physically-distinct disorder-driven contributions to the photocurrents~\cite{KrautBaltz_1979,Azpiroz_2018,Sodemann_2019,Morimoto_2016,Nagaosa_2020,dgoGuest_2021,asgari_2021}. According to our calculations, which do not assume any approximations on the nature of photo-induced electronic processes,  in a large range of $\Gamma$ beyond 300\,meV the photocurrents in FGT exhibit a clear $1/\Gamma$ behavior, see inset of Fig.~\ref{fig:FGT_001_linPol_nk8k}. On the other hand the behavior found for smaller disorder with respect to $\Gamma$ appears to be very non-linear with higher-order contributions clearly at play.

In order to examine whether the magnitude of the photocurrents can be controlled by the frequency of the light, we compute relevant components of the conductivity tensor as given by Eq.~\ref{eq:symmetry}, presenting the results in Fig.~3(a). The calculated signal exhibits a very strong variation with frequency, although the qualitative behavior of the two components is similar. 
The observed strong variation can be attributed to the complex orbital composition of the electronic structure of FGT. In particular, pronounced variations just below the frequency of 1\,eV and 2\,eV can be attributed to the transitions between the groups of bands just above and below the Fermi energy at the K-point, and the groups positioned at approximately $-1$\,eV and $+$0.5\,eV at the $\Gamma$-point $-$ which also mediate the behavior of the magneto-optical conductivity of FGT~\cite{arXiv:2108.02926,arXiv:2012.04285v1}. Concerning the overall magnitude of the computed signal, reaching as much as 200\,$\mu$A/V$^2$ for frequencies below 1\,eV, it is comparable to previously reported results in van der Waals monolayers such as non-magnetic GeS and WS$_2$~\cite{Azpiroz_2018,Wang_2017} and more recently in 2D magnets like CrI$_3$~\cite{Zhang_2019}.   

\begin{figure}[t!]
\begin{center}
\rotatebox{0}{\includegraphics [width=0.95\columnwidth]{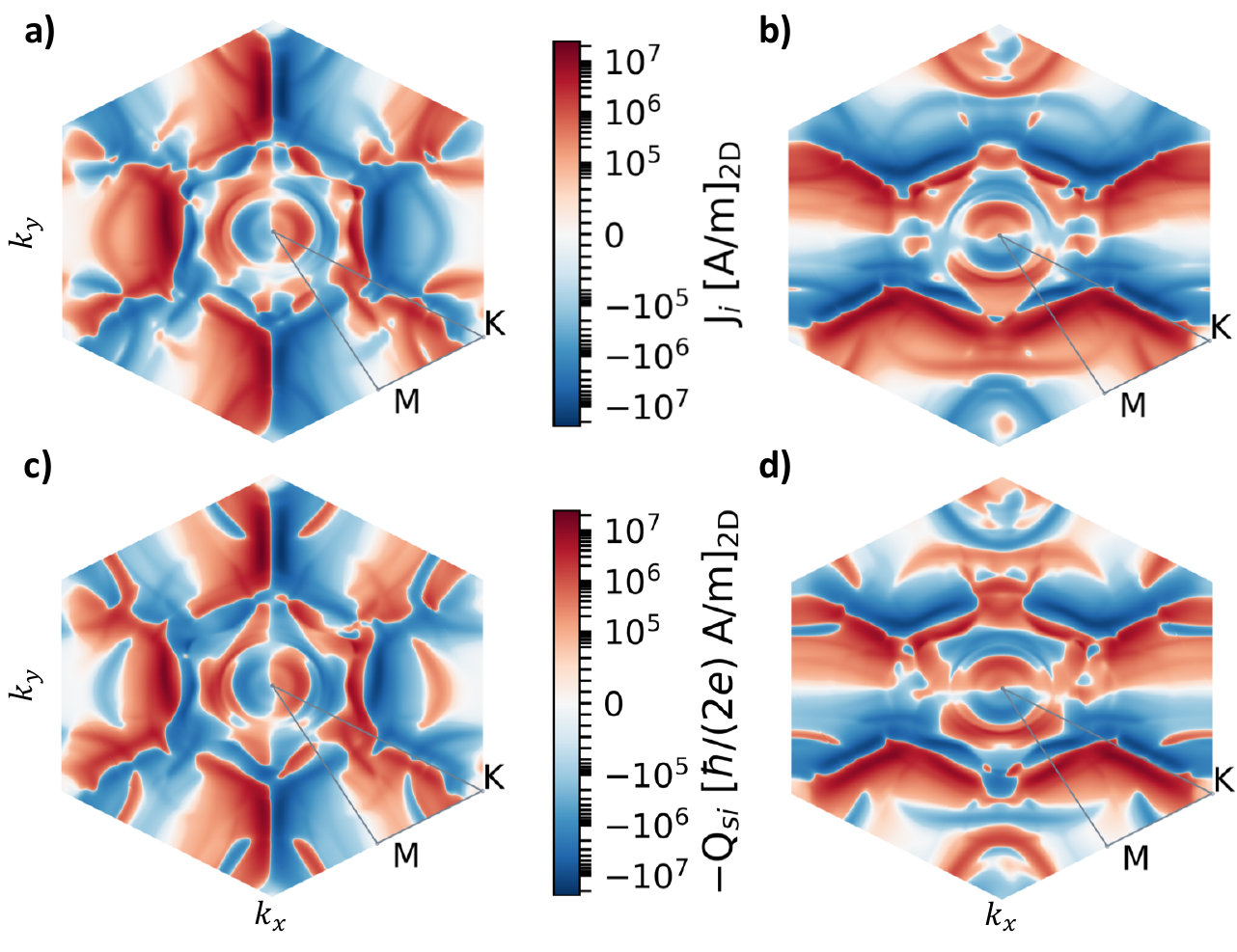}}
\end{center}
\caption{Reciprocal-space distribution of photocurrents. Distribution of the $J_x$ (a) and $J_y$ (b) charge, and $-Q_{zx}$ (c) and $-Q_{zy}$ (d) spin photocurrents in the Brillouin zone of  single-layer FGT computed for the true position of the Fermi energy. The light pulse of frequency $\hbar \omega=1.55$\,eV and intensity  of 10\,GW/cm$^2$
polarized along the $x$-axis was used in combination with disorder strength $\Gamma$  of 25\,meV.  
}
\label{fig:FGT_001_110_pol}
\end{figure}

The signatures of the orbital structure of FGT can be also clearly seen in the dependence of the photocurrents on band filling, shown in Fig.~3(b). For both components of the currents, the suppression of the signal can be clearly visible for a position of the Fermi energy at about $-0.5$\,eV and $-1.5$\,eV, which corresponds to the suppression of corresponding transitions, discussed above, as the states become unpopulated with decreasing band filling. Notably, in contrast to $J_y$, at an energy of $\approx -1.5$\,eV the sign of $J_x$ changes, which should result in a drastic change in the direction of the in-plane photocurrent.   
In Fig.~3(c) we plot the band filling dependence of $Q_{zx}$ and $Q_{zy}$ components of  the  photocurrents of spin. The predicted magnitude of the spin currents which can be generate by light in FGT is sizeable, which marks FGT as one of the promising materials for spin photogalvanic applications. 

Interestingly, in a wide energy window just below  the true $E_F$ the qualitative behavior of the spin currents is very similar to that of charge currents in terms of the evolution of the magnitude with band filling and dependence on disorder strength.   
However, one difference is that while $Q_{zx}$ changes sign around $-1.5$\,eV in correlation with $J_x$, the sign change occurs also for $Q_{zy}$ but not for $J_y$. This stands in sharp contrast to a simple picture of a spin current being proportional to the spin polarization of the participating states along $z$, scaled by the corresponding component of the charge photocurrent: indeed, since just below the energy of $-1.5$\,eV the spin-polarization of Fe states changes  sign, this consideration explains the change in sign of $Q_{zx}$, however, it fails to mimic the change of sign in  $Q_{zy}$.

To understand this behavior better, we plot 
the reciprocal space distribution of charge and spin currents at the true Fermi energy in Fig.~4. While from these plots it is clear that an overall correlation between two types of currents is present, large parts of the Brillouin zone where charge and spin currents have a reversed sign correlation, are also visible, as can be most prominently seen for $y$-components of the currents around the corners of the Brillouin zone. With decreasing band filling the contribution of the corresponding parts becomes promoted, which can explain the change of the sign of $Q_{zy}$ with respect to $J_y$.
Such non-trivial correlation between the magnitude and direction of charge and spin currents can be used to engineer a desired charge and spin transport setup by band filling.
We also observe that the decay of the spin photocurrents with disorder strength can depend strongly on the band filling: e.g. while both components of the spin currents decay rapidly with $\Gamma$ around the true Fermi energy, the $Q_{zx}$ component displays a stronger disorder robustness in the energy region between  $-3$\,eV and $-2$\,eV. This indicates that the direction of spin photocurrents in FGT can be tuned not only by band filling but also by the degree of disorder. 

To summarize, in this work we used an {\it ab-initio} implementation of Keldysh formalism for second order response to address the properties of charge and spin photocurrents in a single-layer Fe$_3$GeTe$_2$. Our predictions concerning the symmetry of the currents are in full agreement with direct first principles calculations, which predict the magnitude of the photocurrents emerging in FGT to be comparable to those of a magnetic Rashba model with strong spin-orbit interaction. The predicted non-trivial response of the currents to such effects as disorder strength, band filling and frequency marks FGT as a promising platform for crafting the desired properties of the photocurrents, which might prove to be important for future optospintronics applications of 2D magnets.

{\it Acknowledgements.}
This work was supported by the Deutsche Forschungsgemeinschaft (DFG, German Research Foundation) $-$ TRR 173 $-$ 268565370 (projects A11 and A01), TRR 288 – 422213477 (project A06), and the Sino-German research project DISTOMAT (MO 1731/10-1). This project has received funding from the European Union’s Horizon 2020 research and innovation programme under the Marie Skłodowska-Curie grant agreement No 861300, and under synergy grant "3D MAGiC” agreement (Grant No. 856538).
We  also gratefully acknowledge the J\"ulich Supercomputing Centre and RWTH Aachen University for providing computational resources under project  jiff40.


\hbadness=99999 
\bibliographystyle{apsrev4-2}
\bibliography{literature}




\end{document}